\documentclass[fleqn,twoside,twocolumn,nofootinbib]{revtex4} 
\usepackage{ujp} 
\usepackage[cp1251]{inputenc}
\usepackage[ukrainian,english,russian]{babel}
\begin{document}
\title[EQUATION OF STATE FOR WATER IN THE SMALL COMPRESSIBILITY REGION]
{EQUATION OF STATE FOR WATER IN THE SMALL COMPRESSIBILITY REGION}
\author{V.Yu. Bardik}
\affiliation{Taras Shevchenko National University of Kyiv, Physics Faculty}
\address{4, Prosp. Academician Glushkov, Kyiv 03127, Ukraine}
\email{vital@univ.kiev.ua, pavlov@univ.kiev.ua}
\author{D. Nerukh}%
\affiliation{Aston University, Non-linearity and Complexity Research Group}
\address{Birmingham, B4 7ET, UK}
\author{E.V. Pavlov}%
\affiliation{Taras Shevchenko National University of Kyiv, Physics Faculty}
\address{4, Prosp. Academician Glushkov, Kyiv 03127, Ukraine}
\email{pavlov@univ.kiev.ua}
\author{I.V.~Zhyganiuk}%
\affiliation{I.I. Mechnikov Odessa National University}%
\address{2, Dvoryanskaya Str., Odessa 65082, Ukraine}%
\email{ivz@ukrpost.ua}

\udk{532} \pacs{66.10.C} \razd{\secvi}

\setcounter{page}{47}%
\maketitle

\begin{abstract}
The equation of state for dense fluids has been derived within the framework of the Sutherland and
Katz potential models. The equation quantitatively agrees with experimental data
on the isothermal compression of water under extrapolation into the high pressure region. It establishes
an explicit relationship between the thermodynamic experimental data and the effective parameters
of the molecular potential.
\end{abstract}

\section{Introduction}

Deriving the equation of state (EoS) for water in a wide range
of pressures and temperatures remains a challenging open problem,
especially in the high pressure region. In the recent
papers \cite{1,2,3,4} devoted to the water EoS, this problem was
solved by fitting multiparameter formulae with such a large number
of adjustable parameters that it approaches the number of
experimental points. These methods are not based on reliable
statistical mechanic foundations, and the applicability of these EoS
is restricted. If the functional form of the EoS and their
parameters are applicable to other substance or solution is an open
question.

In comparison with the majority of one-component liquids, water
reveals many unusual properties in its normal and supercooled
states. The analysis of the diffusion peak of the quasi-elastic
incoherent neutron scattering and the kinematic shear viscosity of
water has shown that the global H-bond network disintegrates into an
ensemble of weakly interacting clusters: dimers, trimers, tetramers,
{\it etc.} \cite{5,6,7,8,9}. It was also shown~\cite{10} that the properties of water in the
supercritical region are determined by the
averaged spherically symmetric potential. Therefore, it is
reasonable to use such
well-known models as Lennard-Jones, Buckingham, Sutherland or Katz
potentials for water in the high pressure region.

In  this paper, we derive the EoS for a supercritical fluid within
the framework of the Sutherland and Katz potentials. In our previous
work~\cite{11}, we used a new version of the thermodynamic
perturbation theory (TPT) originally proposed by Sysoev~\cite{11}.
The main feature of the proposed TPT is in the assumption that the
functional form of the perturbed potential is identical to the
potential of the reference system. Therefore, the deviation of
the potential of the more compressed system from the potential of
the less compressed system is considered as a perturbation. On
this basis, the concept of a reference thermodynamic state has been
developed. A functional expansion of the free energy gave the
possibility to derive, at a certain choice of the parameter
expansion, two EoS modifications within the framework of a realistic model
and the ``soft'' sphere potential one. As was shown in~\cite{11},
these EoS correctly described the isothermal compression for
supercritical fluids of inert gases.

Following~\cite{11}, we use the free energy perturbation expansion
\[
F_V \left( {V_0 } \right) = F_0 \left( {V_0 } \right) +
\sum\limits_{i < j} {\int {d\bar r_i } } d\bar r \, \Delta e\left(
{\bar r_{ij} } \right) \, \frac{{\delta F\left( {V_0 }
\right)}}{{\delta e\left( {\bar r_{ij} } \right)}}+
\]
\[
+ \frac{1}{{2!}}\sum\limits_{\scriptstyle i < j \hfill \atop
\scriptstyle e < m \hfill} {\int {d\bar r_i } } d\bar r_j d\bar r_l
d\bar r_m  \, \Delta e\left( {r_{ij} } \right) \, \Delta e\left(
{r_{lm} } \right) \times
\]
\begin{equation}
\times \frac{{\delta ^2 F\left( {V_0 } \right)}}{{\delta e\left( {r_{ij} } \right)\delta e\left( {r_{lm} } \right)}},
\end{equation}
where $\Delta e\left( {r_{ij} } \right) = e\left( {r_{ij} }
\right) - e_0 \left( {r_{ij} } \right) = e^{ - \frac{{\varphi \left(
{qr_{ij} } \right)}}{{kT}}}  - e^{ - \frac{{\varphi \left( {r_{ij} }
\right)}}{{kT}}} $, $\phi \left( {qr_{ij} } \right)$ is the
potential of the perturbed system, and $q$ is a scale factor, $\bar r
\to q\bar r,q = \sqrt[{^3 }]{{\frac{V}{{V_0 }}}}$.

Equation (1) can be transformed into the expression for pressure (the details are given in~\cite{11})
\[
P - P_0  = \frac{{NkT\Delta V}}{V} -
\]
\[
- \frac{{\rho _0 ^2 V_0 }}{{6V}}\int\limits_V {g_2 \left( r \right)}
\exp \left[ {\beta \left( { - \Phi \left( {qr} \right) + \Phi \left(
r \right)} \right)} \right] \, \frac{{d\Phi \left( {qr}
\right)}}{{dr}}rd\bar r +
\]
\begin{equation}
+ \frac{{\rho _0 ^2 }} {6}\int\limits_V {g_2 \left( r \right)}
\frac{{d\Phi \left( {qr} \right)}}{{dr}}rd\bar r.
\end{equation}

This  expression was obtained within the framework of a realistic
potential model, that can be presented in the general form
\begin{equation}
\varphi \left( r \right) = \Phi \left( r \right) + \psi \left( r
\right),
\end{equation}
where $\Phi \left( r \right)$ is the repulsive part of the potential, and
$\psi \left( r \right)$  is its attractive part.

\section{The Equation of State within the Framework of Sutherland and Katz Potential Models}

In  the case of short-range potentials, the expression for the pressure
can be rewritten in the form
\[
P = \frac{{NkT}}{V} - \frac{{\rho _0 ^2 V_0 }}{{6V}}\int\limits_V {g_2 \left( r \right)f\left( r \right)} \upsilon \left( r \right)d\bar r -
\]
\begin{equation}
- \frac{{\rho _0 ^2 V_0 }}{{6V}}\int\limits_V {g_2 \left( r \right)} \upsilon _0 \left( r \right)d\bar r ,
\end{equation}
where  $f\left( r \right) = e^{^{\left[ {\beta \left( { - \varphi
\left( {qr} \right) - \varphi \left( r \right)} \right)} \right]} }
- 1$, $\upsilon \left( r \right) =  - \bar \nabla \varphi \left(
{qr} \right)\bar r$ is the virial of intermolecular forces and
$\upsilon _0 \left( r \right) =  - \bar \nabla \varphi \left( {qr}
\right)\bar r\left| {_{q = 1} } \right. =  - \bar \nabla \varphi
\left( r \right)\bar r$ is  the virial of intermolecular forces in
the reference state. With regard for the expression for the pressure
in the reference state
\begin{equation}
P_0  = \frac{{NkT}}{{V_0 }} - \frac{{\rho _0 ^2 }}{6}\int\limits_V {g_2 \left( r \right)} \upsilon _0 \left( r \right)d\bar r,
\end{equation}
we rewrite relation (4) as
\[
P = \frac{{V_0 P{}_0}}{V} - \frac{{\rho _0 ^2 V_0 }}{{6V}}\int\limits_V {g_2 \left( r \right)f\left( r \right)} \upsilon \left( r \right)d\bar r -
\]
\begin{equation}
 - \frac{{\rho _0 ^2  V_0 }}{{6V}}\int\limits_V {g_2 \left( r \right)\Delta } \upsilon \left( r \right)d\bar
 r.
\end{equation}

Equation (6) can be expressed in the terms of $\Pi  = \frac{{P - P_0
}}{{P_0 }}$ or $ \Delta  = \frac{{V_0  - V}}{{V_0 }}$, using of the
approximate quality $\frac{{V_0 }}{V} \approx 1 + \Delta$
\begin{equation}
\Pi  = \Delta  + \left( {1 + \Delta } \right)\left( {L + K} \right),
\end{equation}
where
\begin{equation}
L =  - \frac{{\rho _0^2 }}{{6P_0 }}\int\limits_V {g_2 \left( r \right)f\left( r \right)} \upsilon \left( r \right)d\bar r ,
\end{equation}
\begin{equation}
K =  - \frac{{\rho _0^2 }}{{6P_0 }}\int\limits_V {g_2 \left( r \right)} \Delta \upsilon \left( r \right)d\bar r .
\end{equation}

Now  the problem of deriving the EoS reduces to the evaluation of
integrals (8) and (9). First, we consider the Sutherland potential
\begin{equation}
\varphi \left( r \right) = \left\{ {\begin{array}{ll}
   {\infty ,} & {r < d_0, }  \\
   { - cr^{ - m} ,} & {r > d_0, }  \\
\end{array}} \right.
\end{equation}
$d_0$  is the molecular diameter, and the potential well depth
$\varepsilon$ is defined by $\varepsilon  = \frac{c}{{d_0^m }}$.
Then $\varphi \left( {qr} \right)$ is written as
\begin{equation}
\varphi \left( {qr} \right) = \left\{ {\begin{array}{ll}
   {\infty ,} & {r < d_0, }  \\
   { - c\left( {\frac{{V_0 }}{V}} \right)^{m/3} r^{ - m} ,} & {r > d_0, }  \\
\end{array}} \right.
\end{equation}
where  $d = \sqrt[3]{{\frac{{V_0 }}{V}}}d_0$. We calculate the
singular force $ - \frac{{d\varphi \left( r \right)}}{{dr}}$
following~\cite{13} and obtain
\begin{equation}
\upsilon \left( {\bar r} \right) =  - kT\delta \left( {r - d}
\right)r + cm\left( {\frac{{V_0 }}{V}} \right)^{m/3} r^{ - m} \Theta
\left( {r - d} \right) ,
\end{equation}
\[
\Delta \upsilon \left( {\bar r} \right) =   - kT\left[ {\delta
\left( {r - d} \right) - \delta \left( {r - d_0 } \right)} \right]r
+
\]
\begin{equation}
 + cmr^{ - m} \left[ {\left( {\frac{{V_0 }}{V}} \right)^{m/3} \Theta \left( {r - d} \right) - \Theta \left( {r - d_0 } \right)} \right] ,
\end{equation}
$\delta \left( {r - d} \right)$ is the Dirac delta function and
$\Theta \left( {r - d} \right)$ is the Heaviside step function. The
expression for $L$ takes the form
\[
L =  - \frac{{2\pi \rho _0^2 }}{{3P_0 }}\bigg[ { - kTg_2 \left( d
\right)f\left( {r \downarrow d} \right)d^3  + }
\]
\begin{equation}
 { + cm\left( {\frac{{V_0 }}{V}} \right)^{m/3}
\int\limits_d^\infty  {g_2 \left( r \right)r^{ - m + 2} f\left( r
\right)} dr} \bigg]
\end{equation}
where $f\left( {r \downarrow d} \right) = \mathop {\lim
}\limits_{r \to d + 0} f\left( r \right)$ \cite{13}. The integral in
(14) can be evaluated if we assume that $g_2 \left( r \right)
\approx e^{ - \beta \varphi \left( r \right)}.$ Then, in view of
$\frac{{V_0 }}{V} \approx 1 + \Delta,$ we have
\[
cm\left( {\frac{{V_0 }}{V}} \right)^{m/3} \int\limits_d^\infty  {g_2
\left( r \right)r^{ - m + 2} f\left( r \right)} d =
\]
\[
 = \left( { - \beta } \right)^{(3-m)/m} c^{3/m} \left[ {\left( {\frac{{V_0 }}{V}} \right)^{2m/3} \gamma \left( {x,y} \right) - } \right.
\]
\begin{equation}
\left. { - \left( {\frac{{V_0 }}{V}} \right)^{m/3} \gamma \left(
{x,y\left( {\frac{V}{{V_0 }}} \right)^{m/3} } \right)} \right] .
\end{equation}

Here, $\gamma \left( {x,y} \right)$ is the upper incomplete gamma
function with $x = \frac{{m - 3}}{\beta }$, $y =  - \frac{{\beta
c}}{{d_0^m }}$. Using the same assumptions for integral (9), we
obtain
\[
K =  \frac{{2\pi \rho _0^2 }}{{3P_0 }}\left[ {kT\left[ {g_2 \left( d
\right)d^3  - g_2 \left( {r \downarrow d_0 } \right)d^3 } \right] -
} \right.
\]
\[
 - \left( { - \beta } \right)^{(3-m)/m} c^{3/m}  \times
\]
\begin{equation}
\times \left. {\left[ {\left( {\frac{{V_0 }}{V}} \right)^{2m/3}
\gamma \left( {x,y\left( {\frac{V}{{V_0 }}} \right)^{m/3} } \right)
- \gamma \left( {x,y} \right)} \right]} \right] ,
\end{equation}
where  $g_2 \left( {r \downarrow d_0 } \right) = \mathop {\lim
}\limits_{r \to d + 0} g_2 \left( r \right)$. Since
the parameter $\Delta$ is small, we can expand $g_2 \left( d
\right)$ at a point $d_0$ in the Taylor series
\begin{equation}
g_2 \left( d \right)  \approx g_2 \left( {r \downarrow d_0 }
\right)\left[ {1 + C\left( T \right)\Delta } \right],
\end{equation}

\begin{table}[b]
\noindent\caption{The values of the EoS
parameters}\vskip3mm\tabcolsep4.5pt \noindent{\footnotesize
\begin{tabular}{c c c c c c c}
\hline%
\multicolumn{1}{c}{\rule{0pt}{9pt}  $T$, K }%
& \multicolumn{1}{|c}{$300$} & \multicolumn{1}{|c}{$350$} &
\multicolumn{1}{|c}{$400$} & \multicolumn{1}{|c}{$450$} &
\multicolumn{1}{|c}{$500$}
& \multicolumn{1}{|c}{$550$} \\
\hline
$B(T)$, MPa    & $9.63$ & $8.55$ & $7.70$ & $6.55$ & $5.10$ & $3.35$ \\
$C(T)$         & $2.76$ & $2.65$ & $2.59$ & $2.45$ & $2.39$ & $2.10$  \\
$A(T)$, K$^{-1}$ & $0.0042$ & $0.0037$ & $0.0032$ & $0.0027$ & $0.0024$ & $0.002$ \\
\hline
\end{tabular}
}
\end{table}

\noindent where  $C(T) = \frac{{d_0 }}{3}\frac{{\partial \ln
g(r)}}{{\partial r}}\left| {_{r \to d_0  + 0} } \right.$ is the
function of the reference state. Since, in the approximation $\Delta
\ll 1,$ $ f\left( {r \downarrow d} \right) = \exp \left(
{\frac{{\beta cm\Delta }}{{3d_0^m }}} \right) - 1$ and $\frac{\Delta
}{\Pi } \ll 1,$ Eq. (7) in the case of small compressibility takes
the form
\[
P = P_0  + \left[ {B\left( T \right) + P_0 } \right]  \bigg[ \left(
{e^{\frac{\Delta }{{A\left( T \right)  T}}} \!\! - 1} \right)\left[
{1 + \Delta \left( {2 + C\left( T \right)} \right)} \right] -
\]
\begin{equation}
  - \left( {1 + C\left( T \right)} \right)\Delta  + 2\left( {1 + C\left( T \right)} \right)\Delta ^2  \bigg] + \Gamma ,
\end{equation}
$\Gamma$  is the term which comprises the incomplete gamma function
\begin{equation}
\Gamma  = \frac{{2\pi \rho _0^2 }}{{3P_0 }}\left[ {\left( { - \beta
} \right)^{(3-m)/m} c^{3/m} \left[ {\left( {\frac{{V_0 }}{V}}
\right)^{2m/3} \!\! - 1} \right]\gamma \left( {x,y} \right)} \right]
.
\end{equation}

For the Sutherland model, it is expressed in terms of the second
virial coefficient ${B_2 \left( T \right)}$
\begin{equation}
\Gamma  = \frac{{kT\rho _0 }}{{P_0 }}\left[ {\left( {\frac{{V_0
}}{V}} \right)^{2m/3}  - 1} \right]\left[ {B_2 \left( T \right) +
\frac{{2\pi d_0^3 }}{3}} \right] .
\end{equation}

However,  the analysis of the experimental data for some dense
fluids (water, argon, neon, krypton) revealed that this term can be
neglected under the condition $\Pi  \approx 10^3$ with an accuracy of 1\%.
It is the fairly wide range of thermodynamic variables, where the
isothermal compressibility is low ($\Delta \ll 1$) corresponding to
the pressure interval 100--2200~MPa. The terms $ 2\left( {1 +
C\left( T \right)} \right)\Delta ^2  - \left( {1 + C\left( T
\right)} \right)\Delta$ can also be ignored with the same accuracy.
Finally, we arrive at the EoS within the framework of the Sutherland
model
\begin{equation}
P = P_0  + \left[ {B\left( T \right) + P_0 } \right] \, \left[
{\left( {e^{\frac{\Delta }{{A\left( T \right) \, T}}}  - 1} \right)
\, D\left( T \right)} \right],
\end{equation}
where $D\left( T \right) = 1 + \Delta \left( {2 + C\left( T
\right)} \right)$. Expression (21) is a three-constant equation
of state with the adjustable parameters $B(T)$, $D(T),$ and $A(T)$ (see Table 1).
The parameter $B(T)$ depends on the temperature. It is related to the
pressure caused by attractive forces
\[
B\left( T \right) = \frac{{2\pi \rho _0^2 }}{{3_0 }}kTg_2 \left( {r \downarrow d_0 } \right)d_0^3  - P_0  =
\]
\begin{equation}
 = P_0^r  - P_0  = P_0^a  - \frac{{NkT}}{{V_0 }} ,
\end{equation}

\begin{figure}
\includegraphics[width=\column]{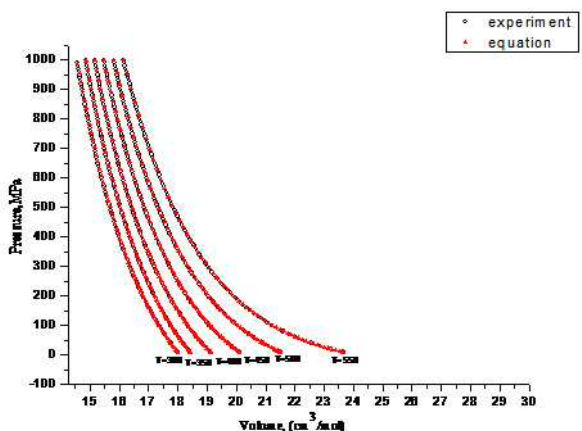}
\vskip-3mm\caption{Processing of the experimental PVT-data for
supercritical water}
\end{figure}

\noindent where $P_0^r$ is the pressure caused by the repulsive
forces in the reference state, and  $P_0^a$ is a part of the pressure
due to the attractive forces. The parameter $A(T)$, within the
framework of the Sutherland model, is expressed by the formula
\begin{equation}
A(T) = \frac{{3 \, d_0 ^m  \, k}}{{mc}}.
\end{equation}

If we use the Katz potential model
\begin{equation}
\Phi \left( {\bar r} \right) = \left\{ {\begin{array}{ll}
   {\infty ,} & {r < d_0, }  \\
   { - \frac{{a R_0^3 }}{{4\pi }}e^{ - r R_0 } ,} & {r > d_0, }  \\
\end{array}} \right.
\end{equation}
where $R_0$ is the interaction radius of the attractive forces,
$\frac{{a R_0^3 }}{{4\pi }}$ is the interaction constant, Eq.(21)
retains its functional form, but the parameter $A(T)$ is defined by
the formula $ A(T) = \frac{{12 \, \pi d_0 ^m \exp \left( {R_0 d_0 }
\right) \, k}}{{d_0 R_0^4 a}} $.

The parameters $B(T)$ and $C(T)$ are defined in a similar way.

\section{Experimental PVT-data Analysis}

We used  the technique from the previous paper~\cite{11} to process
the PVT data and evaluate the EoS parameters. It turns out that
Eq.~(21) yields good agreement with the experimental PVT
data~\cite{14} under the extrapolation to the high pressure region.
The results of the comparison are presented in Figure.

Of  special interest are the values of the parameter $A(T),$ because
it is related to the parameters of the potential, formulae (23) and
(25) for the Sutherland and Katz models. On the basis of (23), we
evaluated the values of the potential well depth $\varepsilon$ at a
fixed value $m =12$ commonly used in the Lennard--Jones model
(Table~2).

\begin{table}[b]
\noindent\caption{The values of \boldmath$\varepsilon$ (Sutherland
potential) at $m=12$ in the temperature interval 300--550~K
}\vskip3mm\tabcolsep5.5pt \noindent{\footnotesize \begin{tabular}{c
c c c c c c}
 \hline%
 \multicolumn{1}{c}{\rule{0pt}{9pt} $T$, K  }%
 & \multicolumn{1}{|c}{$300$}
 & \multicolumn{1}{|c}{$350$}
 & \multicolumn{1}{|c}{$400$}
 & \multicolumn{1}{|c}{$450$}
 & \multicolumn{1}{|c}{$500$}
 & \multicolumn{1}{|c}{$550$} \\
 \hline
$\varepsilon$,  $J 10^3$/mole & $0.493$ & $0.562$ & $0.650$ & $0.769$ & $0.866$ & $1.038$ \\
 \hline
 \end{tabular}
}
\end{table}

If  we fix the value of $\varepsilon = 0.650$  kJ/mole, which
corresponds to the SPC/E model, we obtain a variation of the
softness parameter $m$ with the temperature. In general, fixing
$\varepsilon$ at the values for well-known potential models such as
SPC/Fw~\cite{15}, TIP3P/Fw~\cite{15}, and TIP5P/Ew~\cite{16} leads to
a variation of the softness parameter $m$ with the temperature within
the framework of the Sutherland potential.

\section{Conclusion}

The  approach developed on the basis of the free energy perturbation
expansion and a new version of TPT resulted in some universality for
the EOS statistical foundation of the low weight molecular
supercritical fluids. In principle, the concept of the thermodynamic
reference state implies that an initial state ($P_0$, $V_0$) on the
isotherm corresponds to the reference system with the unperturbed
potential, and every subsequent point on the isotherm ($P_1$, $V_2$),
\dots, ($P_n$, $V_n$) corresponds to the system with the perturbed
potential at the isothermal compression of the system. This modification
of TPT allowed obtaining the EoS which exhibits good results under
the extrapolation to the high-pressure region and, most importantly,
establishes a relationship between the parameters of the model
potential and the thermodynamic properties of substances. This
relationship gives estimations for the values of the parameter
$\varepsilon$ (Table 2). Interestingly, the values are of the same order
of magnitude as the values of many well-known water models (SPC/E,
SPC/Fw, TIP3P/Fw, TIP5P/Ew). However, the temperature dependence of
$\varepsilon$ in this region of thermodynamic variables indicates
that the form of the Sutherland model is unsuitable for this high
pressure region. Nevertheless, these data can be used as additional
information for calibrating the potential parameters in
simulations.

\rezume{%
РІВНЯННЯ СТАНУ ВОДИ В ОБЛАСТІ МАЛИХ\\ СТИСЛИВОСТЕЙ}{В.Ю. Бардік,
Д.А. Нерух,  Є.В. Павлов, І.В. Жиганюк} {Одержано статистично
обґрунтоване рівняння стану густих флюїдів у рамках моделей
потенціалу Сюзерленда та Каца.  Запропоноване рівняння стану з
високою точністю  узгоджується з експериментальними даними по
ізотермічному стисненню води при екстраполяції в область високих
тисків. Встановлено кількісний зв'язок між ефективними параметрами
модельних потенціалів з параметрами рівняння стану.}


\begin{thebibliography}{16}

\bibitem {1}A.~Saul and W.~Wagner, J.~Phys.~Chem. Ref. Data \textbf{18},  1537 (1989).\vskip3mm

\bibitem {2}W.~Wagner and A.~Pruss, J.~Phys.~Chem. Ref. Data \textbf{31},  387
(2002).\vskip3mm

\bibitem {3}R.I.~Nigmatulin and R.Kh.~Bolotnova, Teplofiz. Vys. Temp. \textbf{46},  182  (2008).\vskip3mm

\bibitem {4}R.I.~Nigmatulin and R.Kh.~Bolotnova, Teplofiz. Vys. Temp. \textbf{46},  325  (2008).\vskip3mm

\bibitem {5}A.I.~Fisenko and N.P.~Malomuzh, Int. J. Mol. Sc. \textbf{10}, 2383  (2009).\vskip3mm

\bibitem {6}A.I.~Fisenko and N.P.~Malomuzh, Chem.~Phys. \textbf{345}, 164  (2008).\vskip3mm

\bibitem {7}L.A.~Bulavin, A.I.~Fisenko, and N.P.~Malomuzh, Chem. Phys. Lett. \textbf{453}, 183  (2008).\vskip3mm

\bibitem {8}A.I.~Fisenko, N.P.~Malomuzh, and A.V.~Oleynik, Chem. Phys. Lett. \textbf{450}, 297  (2008).\vskip3mm

\bibitem {9}T.V.~Lokotosh, N.P.~Malomuzh, and K.N.~Pankratov, J.~Chem. Eng. Data \textbf{55}, 2021 (2010).\vskip3mm

\bibitem {10}S.V.~Lishchuk, N.P.~Malomuzh, and P.V.~Makhlaichuk, Phys. Lett.~A \textbf{374}, 2084 (2010).\vskip3mm

\bibitem {11}V.Yu. Bardik, N.P. Malomuzh, K.S. Shakun, and V.M.~Sy\-soev, Journal of Molecular Liquids, in press.\vskip3mm

\bibitem {12}V.M.~Sysoev, Teor. Mat Fiz. \textbf{55}, 305 (1983).\vskip3mm

\bibitem {13}R.~Reijnhart, Physica A \textbf{83}, 533 (1976).\vskip3mm

\bibitem {14}NIST Database http://webbook.nist.gov/chemistry/fluid/.\vskip3mm

\bibitem {15}Y.~Wu, H.L.~Tepper, G.A. Voth, J.~Chem. Phys. \textbf{124}, 024503 (2006).\vskip3mm

\bibitem {16}S.W.~Rick, J.~Chem. Phys. \textbf{120}, 6085 (2004).\vskip3mm
\begin{flushright}
{\footnotesize  Received 17.05.11}
\end{flushright}
\end{thebibliography}
\end{document}